%% file: main.tex
\documentclass[%
 reprint,
amsmath,amssymb,
aps,
pra,
]{revtex4-1}

\usepackage{graphicx}
\usepackage{dcolumn}
\usepackage{bm}
\usepackage[mathlines]{lineno}
\usepackage{amsmath}
\usepackage{relsize}%
\usepackage{verbatim}

\begin{document}

\preprint{APS/123-QED}
\title{Theoretical study on memory-based optical converter with degenerate Zeeman states}

\author{Pin-Ju Tsai$^{1,2}$}
\author{Yan-Cheng Wei$^{1,2}$}
\author{Bo-Han Wu$^{2}$}
\author{Sheng-Xiang Lin$^{1,2}$}
\author{Ying-Cheng Chen$^{2,3}$}

\email{Corresponding author chenyc@pub.iams.sinica.edu.tw}

\affiliation{$^{1}$Department of Physics, National Taiwan University, Taipei 10617, Taiwan}
\affiliation{$^{2}$Institute of Atomic and Molecular Sciences, Academia Sinica, Taipei 10617, Taiwan}
\affiliation{$^{3}$Center for Quantum Technology, Hsinchu 30013, Taiwan}

\date{\today}
\begin{abstract}
We present a theoretical study on the efficiency variation of coherent light conversion based on optical memories using the electromagnetically induced transparency (EIT) protocol in an atomic system with degenerate Zeeman states. Based on the Maxwell-Bloch equation, we obtain an approximate analytic solution for the converted light pulses which clarifies that two major factors affecting the efficiency of the converted pulses. The first one is the finite bandwidth effect of the pulses and the difference in the delay-bandwidth product of the writing and reading channel due to the difference in the transition dipole moment. The second one is the mismatch between the stored ground-state coherence and the ratio of the Clebsch-Gordan coefficient for the probe and control transition in the reading channel which results in a non-adiabatic energy loss. To correspond to real experimental conditions, we also perform a numerical calculation of the variation in conversion efficiency versus the Zeeman population distribution under the Zeeman-state optical pumping in storing a $\sigma^{+}$-polarized pulse and retrieving with $\sigma^{-}$ polarization in cesium atoms. Our work provides essential physical insights and quantitative knowledge for the development of a coherent optical converter based on EIT-memory.

\begin{description}
\item[PACS numbers]
32.80.Qk, 42.50.Gy
\end{description}
\end{abstract}

\pacs{Valid PACS appear here}
\maketitle

\section{\label{sec:level1}Introduction}
Electromagnetically induced transparency (EIT) and the associated slow light effect in a $\Lambda$-type three-level system offers an avenue for the implementation of optical quantum memories which have numerous applications in quantum information processing\cite{PhysRevLett.84.5094,PhysRevA.65.022314}. By adiabatic ramping off the control field, the coherence among the atomic ground states generated by a weak probe pulse and the control field can be written and stored inside the atomic medium. After a certain storage time, the control field is turned on to beat with the atomic coherence and the written probe information is retrieved as an output optical pulse\cite{Nature.409.490,PhysRevLett.86.783}. The temporal width, frequency, and propagation direction of the retrieved probe pulse can be manipulated by varying the intensity, frequency, and propagation direction of the control field during the reading process.\cite{PhysRevLett.88.103601,PhysRevA.69.035803,PhysRevA.72.053803,OptLett.31.3511,OptLett.32.2771}. By adding a fourth active excited state to form a four-level double-$\Lambda$ system one can store the probe pulse with one $\Lambda$ system and turn on the second control field to release the optical pulse in the other $\Lambda$ system such that either its frequency is far away from the probe pulse or its polarization is different. These properties can be used to implement a coherent optical converter in a quantum network bridging different quantum devices\cite{Nature.453.1023}, each of which only interacts with light of specific properties. Furthermore, one can turn on the control fields of both $\Lambda$ systems simultaneously during the reading process in order to retrieve the two optical pulses with different frequency components\cite{OptCommun.209.149,OptCommun.217.275,PhysRevA.69.043801,PhysLettA.346.269,PhysRevA.73.013804,PhysRevA.77.013823}. The amplitudes of the two frequency components can be tund by varying the intensity ratio of the two control fields. Such a system can be used as a controllable frequency beam splitter for photons\cite{PhysLettA.346.269,PhysRevA.72.043801,OptLett.32.2771,SciRep.6.34279,PhysRevA.97.063801,PhysRevA.97.063805}.   

However, some complications are unavoidable when implementing the coherent optical converter or frequency beam splitter with a double-$\Lambda$ system in real atoms since each atomic state usually contains Zeeman degeneracy. In the memory-based optical converter,  we define the conversion efficiency as the ratio of the energy of the retrieved pulse in the second channel to that of the input probe pulse. Quantitative knowledge of the conversion process and conversion efficiency under such realistic situations is important and is helpful in all related experiments. However, such studies have been rare, with the exception of  Refs.\cite{OptLett.31.3511,PhysRevA.75.013812} in which the authors examined the role of degenerate Zeeman states in polarization conversion using EIT memories in the adiabatic limit. They showed that it is the incompatibility between the stored ground-state coherence and the ratio of the probe and control Clebsch-Gordan coefficients in the reading channel which is responsible for some energy loss during the reading process. For brevity, We call this the coherence mismatch factor. They derived a formula for the conversion efficiency in the adiabatic limit, which is related to the Zeeman population distribution and the Clebsch-Gordan coefficients of all transitions involved. Based on this, they pointed out that such an energy loss could be avoided if the population is prepared in a single Zeeman state. 

This work extends that study to go beyond the adiabatic condition. This treatment is important, since in realistic situations, optical pulses are used in memory-based conversion where the finite bandwidth effect needs to be taken into consideration. Based on the the Maxwell-Bloch equation, we derive an approximate analytic formula for the conversion efficiency. In addition to the coherence mismatch factor which affects the conversion efficiency, our results show that the finite bandwidth effect and the difference in the transition dipole moment between the writing and reading channels is also  important. Even if the whole population is prepared in a single Zeeman state, this finite-bandwidth factor in the conversion efficiency may be different than unity, depending upon the delay-bandwidth product of the reading and writing channels and the ratio ($\eta$) of the group delay time ($T_d$) to the input pulse FWHM duration ($T_p$), $\eta\equiv\frac{T_d}{T_p}$, during the storage process. In the adiabatic limit, this factor approaches unity and the results are the same as that of Ref.\cite{PhysRevA.75.013812}. 

In order to act as a guide to the realistic experiments, we also perform a numerical calculation of the conversion efficiency versus the Zeeman population distribution in an example of polarization converter in cesium atoms under two different optical pumping cases. In the case where a circularly-polarized optical pumping beam is used to pump the Zeeman population towards the outermost Zeeman state, both factors affect the conversion efficiency. In the ideal case with the entire population in the outermost Zeeman state, only the finite-bandwidth factor affects the conversion efficiency. In the case with a $\pi$-polarized optical pumping beam, the population will be pumped towards the $|m=0\rangle$ Zeeman state and distributed symmetrically in the Zeeman manifold with respect to the $|m=0\rangle$ state. Due to this symmetric population distribution and the symmetric Clebsch-Gordon coefficients for the $\sigma^{+}$ and $\sigma^{-}$ transition, the finite-bandwidth factor does not affect the conversion efficiency, only the ground-state coherence mismatch factor does so. In the ideal case with the entire population in the $|m=0\rangle$ state, neither factors affects the conversion efficiency. The analyses provide essential physical insights and quantitative knowledge useful for understanding coherent light conversion based on EIT memories.                       

The rest of this paper is structured as follows: In Sec.\ref{sec0}, which include three subsections, we consider a general case for an optical converter based on EIT-memory in an atomic system with M sets of $\Lambda$-type sub-systems. In Sec.\ref{sec1}, we derive a general relation for the stored ground-state coherence in the writing process, as well as an approximate analytic expression with an input pulse with a Gaussian waveform. In Sec.\ref{sec2}, we derive an expression for the retrieved optical pulse in the converted channel with a given stored ground-state coherence. In Sec.\ref{sec3}, we analyze the characteristics of the converted optical field. In Sec.\ref{sec4} and Sec.\ref{sec5}, we consider a specific case of polarization conversion with the population in a single Zeeman state and in multiple Zeeman states, respectively. In Sec.\ref{sec5}, we discuss the numerical simulation of the conversion efficiency versus the dispersed Zeeman population in a cesium $D_1$-line system under the optical pumping condition. Finally, we conclude this work in Sec.\ref{sec6}. The calculation of the Zeeman-state optical pumping process used in Sec.\ref{sec5} is described in detail in Appendix.

\begin{figure*}[th]
\centering
\includegraphics[width=1\textwidth]{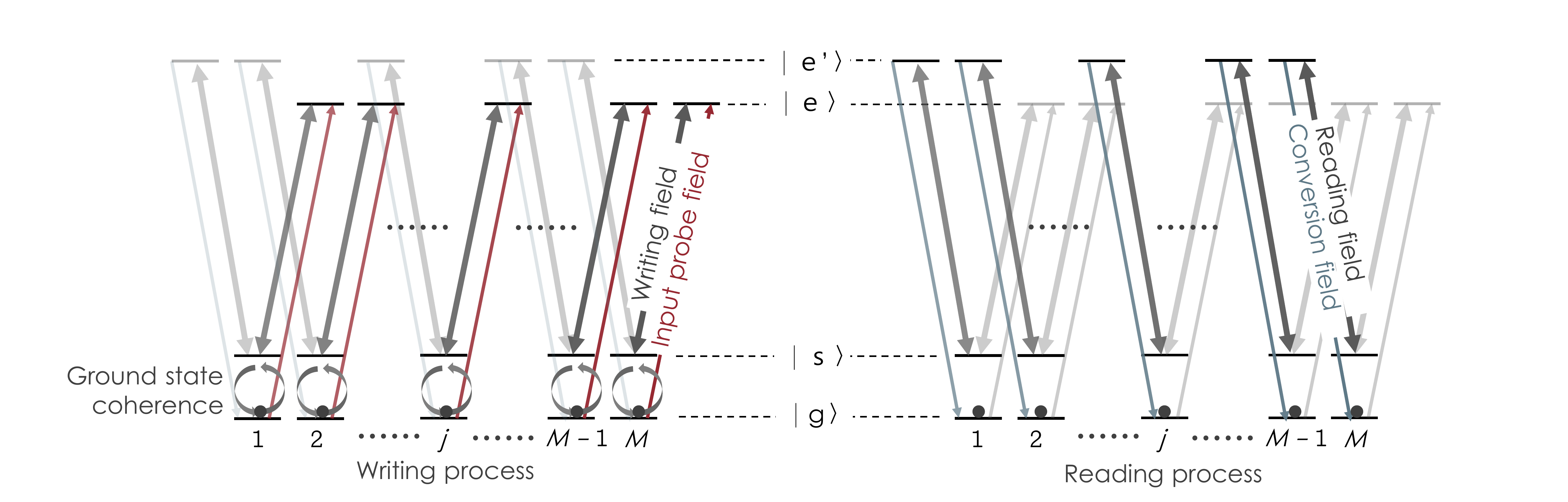}
\caption{Energy level diagram and relevant laser transitions of the memory-based optical converter. This system includes $2M$ ground states and $2M$ excited states. The input probe field $\mathcal{E}_p$ and the strong writing field build up $M$-sets of EIT subsystem in which the probe field is stored and converted to the ground state coherence during the writing process. During the reading process, the reading field drives another transition and convert the coherence to a new field: the converted field $\mathcal{E}_c$.}
\label{process}
\end{figure*}

\section{\label{sec0}Optical converter based on EIT-memory}
We consider the EIT-memory-based optical converter in a multi-level atomic system shown in Fig.\ref{process}. A $\Lambda$-type EIT system is formed with a weak probe field $\mathcal{E}_p(z,t)$ driving the $|g\rangle\rightarrow|e\rangle$ transition and a strong writing field with a Rabi frequency of $\Omega_w$ driving the $|s\rangle\rightarrow|e\rangle$ transition. The memory-based optical converter has three phases. In the writing phase, the information for the weak probe field is written into the collective ground-state coherence (also called the spin wave) of the atomic ensembles by turning off the writing field at time $t=t_w$. In the storage phase, the collective atomic coherence is stored for a time period of $t_s$. For simplicity, we assume that the stored coherence is perfectly maintained during this period so it is not necessary to discuss this phase. In the reading phase, a strong reading field which drives the $|s\rangle\rightarrow|e'\rangle$ transition with a Rabi frequency of $\Omega_r$ is turned on at time $t=t_w+t_s$. The spin wave is then converted into an optical field $\mathcal{E}_c(z,t)$ at the $|g\rangle\rightarrow|e'\rangle$ transition. The memory-based optical conversion is highly related to the forward resonant-type four-wave mixing but the conversion efficiency is not limited to 25\% as in the later case\cite{Opt_Lee,PhysRevA.89.023839,PhysRevA.70.061804,PhysRevA.65.063806}. The energy-level scheme shown in Fig. 1 can be implemented in alkali atoms with $|g\rangle$ and $|s\rangle$ being the two hyperfine ground manifolds with quantum number $F$ and $F+1$, respectively. The two excited states $|e\rangle$ and $|e'\rangle$ could be two different excited hyperfine manifolds with quantum number $F+1$ belonging to two different fine-structure states, or to the same hyperfine manifold but different Zeeman sub-states. 

\subsection{\label{sec1}Writing process}
Under the weak probe field perturbation and the rotating-wave approximation, the optical Bloch equation (OBE) for the relevant atomic coherences of the writing process are:
\begin{equation}
\begin{aligned}
&\frac{\partial}{\partial t}\sigma_{eg,j}=\frac{i}{2}a_{w,j}\Omega_w\sigma_{sg,j}^{(w)}+\frac{i}{2}a_{p,j}p_{j}g_p\mathcal{E}_p-\gamma_{eg,j}\sigma_{eg,j},
\label{si31eq}
\end{aligned}
\end{equation}
\begin{equation}
\frac{\partial}{\partial t}\sigma_{sg,j}^{(w)}=\frac{i}{2}a_{w,j}\Omega_w^*\sigma_{sg,j}^{(w)}-\gamma_{sg,j}\sigma_{sg,j}^{(w)}.
\label{siseeq}
\end{equation}
where $a_{w,j}$ and $a_{p,j}$ are the Clebsch-Gordon coefficient for the writing and probe transition of the $j^{th}$ EIT subsystem, respectively. $p_j$ is the population in the probe ground-state of the $j^{th}$ EIT subsystem. $\gamma_{eg,j}$ is the decay rate of the density matrix element $\sigma_{eg,j}^{(w)}$ and is $\Gamma^w/2$ if spontaneous decay is the dominant relaxation mechanism, where $\Gamma^w$ is the spontaneous decay rate. $\gamma_{sg,j}$ is the decay rate of the ground state coherence. Although we can obtain approximate analytic results for $\gamma_{sg,j}\neq$ 0, the formulae are very cumbersome and the effects of $\gamma_{sg,j}$ are just some additional losses. In order not to obscure the major physics, for simplicity, we assume $\gamma_{sg,j}=$ 0 in our discussion of the analytic formula but include the effects for $\gamma_{sg,j}\neq$ 0 in the numerical calculations. We apply the Fourier transform on the variables $\sigma_{eg,j}(z,t)$, $\sigma_{sg,j}^{(w)}(z,t)$ and $\mathcal{E}_p(z,t)$ in Eqs.(\ref{si31eq})-(\ref{siseeq}) to the frequency domain, e.g. $\mathcal{E}_p(z,\omega)=\frac{1}{\sqrt{2\pi}}\int_{-\infty}^{\infty}e^{i\omega t}\mathcal{E}_p(z,t)$. The frequency-domain OBE reads,
\begin{equation}
\begin{split}
-i\omega\sigma_{eg,j}(z,\omega)=\frac{i}{2}a_{w,j}\Omega_{w}\sigma_{sg,j}^{(w)}(z,\omega)+\frac{i}{2}a_{p,j}p_jg_p\mathcal{E}_p(z,\omega)\\
-\gamma_{eg,j}\sigma_{eg,j}(z,\omega), 
\end{split}
\label{freqOBE1}
\end{equation}
\begin{equation}
-i\omega\sigma_{sg,j}^{(w)}(z,\omega)=\frac{i}{2}a_{w,j}\Omega_w^*\sigma_{eg,j}(z,\omega). 
\label{freqOBE2}
\end{equation}
By solving Eqs.\ref{freqOBE1} and \ref{freqOBE2}, we obtain
\begin{equation}
\sigma_{sg,j}^{(w)}(z,\omega)=A_{w,j}(\omega)p_jg_pR_j^p\frac{\mathcal{E}_p(z,\omega)}{\Omega_w}, 
\label{sig_s}
\end{equation}
\begin{equation}
\sigma_{eg,j}=\frac{-2\omega\sigma_{sg,j}^{(w)}}{a_{w,j}\Omega_w^{*}},
\label{sig_eg}
\end{equation}
where $R^p_{j}=a_{p,j}/a_{w,j}$ and 
\begin{equation}
A_{w,j}(\omega)=-[1-\frac{2i\Gamma_{w}\omega+4\omega^2}{|a_{w,j}\Omega_w|^2}]^{-1}.
\label{Ajw}
\end{equation}
In the case with a slow-varying probe pulse, the solution of the OBE is approximately equal to its steady-state solution and thus one sets $\omega\rightarrow$ 0 in Eq.(\ref{Ajw}). Thus, $A_{w,j}\approx$-1 and from Eq.(\ref{sig_s}) the ground-state coherence is a direct mapping of the probe field in the frequency domain. 

The Maxwell equation for the probe field is 
\begin{equation}
\left(\frac{\partial}{\partial z}+\frac{1}{c}\frac{\partial}{\partial t}\right)\mathcal{E}_p=\frac{ig_pN}{c}\sum_ja_{p,j}\sigma_{eg,j},
\label{MSEp}
\end{equation}
where $g_p=\mu_{eg}\sqrt{\omega_p/2\hbar\epsilon_0V}=\sqrt{\alpha_p\Gamma_wc/2LN}$ is a coupling constant. Performing the Fourier transform on Eq.(\ref{MSEp}) and inserting Eq.(\ref{sig_eg}) into it, we obtain the solution of the frequency-domain probe field as, 
\begin{equation}
\mathcal{E}_p(z,\omega)=\mathcal{E}_p(0,\omega)exp\left(-f^w(\omega)z\right),
\label{slow}
\end{equation}
where $f^w(\omega)$ is,
\begin{equation}
f^w(\omega)=-\frac{i\omega}{c}(1-\frac{2g_c^2N}{|\Omega_w|^2}\sum_jp_j(R_j^p)^2A_{w,}(\omega)).
\end{equation}
At time $t=t_w$, the writing field is turned off to convert the probe field information into the ground-state coherence of the atomic medium. By substituting Eq.(\ref{slow}) into Eq.(\ref{sig_s}) then carrying out the inverse Fourier transform with $t=t_w$, we obtain the stored ground-state coherence of 
\begin{equation}
\sigma_{sg,j}^{(w)}(z)=\frac{p_jg_pR_j^p}{\Omega_w}\mathcal{F}^{-1}[A_{w,j}(\omega)e^{-f^w(\omega)z}\mathcal{E}_p(0,\omega)](t=t_w).
\label{Fsig_s}
\end{equation}
Equation (\ref{Fsig_s}) describes the distribution of the ground-state coherence in space, which is the initial condition of the reading process used in the next subsection. 

As an example for comparison with the experiment, we consider a probe pulse with a Gaussian waveform $\mathcal{E}_p(z=0,t)=\mathcal{E}_0exp(-2ln2(t/T_p)^2)$. The spectral distribution of this pulse is $\mathcal{E}_p(z=0,\omega)=\mathcal{E}_0T_p/\sqrt{4ln2}exp(-(\omega T_p)^2/8ln2)$ with an intensity FWHM bandwidth of $\Delta\omega_0=4ln2/T_p$. By inserting $\mathcal{E}_p(z=0,\omega)$ into Eq.(\ref{Fsig_s}), the exact form of the stored ground-state coherence can be calculated. In order to obtain an approximate analytic formula for the stored coherence, we make some further approximations. We consider Taylor expansion of $A_{w,j}$ and $f^w(\omega)$ in Eq.(\ref{Fsig_s}) with respect to $\omega$, keeping up to the second order term for $f^w(\omega)$. This approximation is valid if $\Delta\omega_0<<min\{\Omega_c^2/\Gamma_w,\Gamma_w\}$. Under this condition, keep only the zero order term for $A_{w,j}$. A closed analytic form for the next order contribution due to O($\omega^1$) of $A_{w,j}$ cannot be obtained, but in that case, one can perform the full numerical calculation based on Eq.(\ref{Fsig_s}) to obtain the exact result. With the above-mentioned approximations, the approximate analytic form of $\sigma^{w}_{sg,j}(z,t_w)$ is:

\begin{equation}
\begin{aligned}
&\sigma^{(w)}_{sg,j}(z,t_w)=\frac{-\mathcal{E}_0p_jg_pa_{p,j}}{a_{w,j}\Omega_w\beta_w(z)}exp\left[-\frac{2ln2(z-v_{w}t_w)^2}{(v_wT_p\beta_w(z))^2}\right],
\label{s_sz}
\end{aligned}
\end{equation}
with the factor of
\begin{equation}
\begin{aligned}
\frac{1}{v_w}&=\frac{1}{c}+\frac{2g_p^2N}{c}\sum_jp_j\frac{a_{p,j}^2}{|a_{w,j}\Omega_w|^2},\\
\end{aligned}
\label{vw}
\end{equation}
\begin{equation}
\begin{aligned}
\frac{1}{\delta\omega_w^2}&=\frac{4g_p^2N L}{cln2}\sum_j\frac{p_ja_{p,j}^2}{|a_{w,j}\Omega_w|^4}\gamma_{eg,j},\\
\end{aligned}
\label{deltaw}
\end{equation}
\begin{equation}
\begin{aligned}
\beta_w(z)&=\left[1+(\frac{4ln2}{T_p\delta\omega_w})^2\frac{z}{L}\right]^{1/2},\\
\end{aligned}
\label{betaw}
\end{equation}
where $v_w$, $\delta\omega_{w}$, and $\beta_w(z)$ represents the group velocity of the probe field, the FWHM EIT transparent
bandwidth and the broadening factor of the probe pulse, respectively and $L$ is the medium length. It should be noted that the group delay time $T_d$ is related to the group velocity by the relation,
\begin{equation}
\begin{aligned}
T_d=L(\frac{1}{v_w}-\frac{1}{c}). 
\end{aligned}
\label{Td}
\end{equation}
In the case with three-level EIT system without Zeeman degeneracy, these relations will be reduced to the simple form shown in Ref.\cite{YF18}.  

Due to the position-dependent pulse broadening factor $\beta_{w}(z)$ in Eq.(\ref{s_sz}), the ground-state coherence cannot maintain the shape of a Gaussian waveform inside the atomic medium in general. However, if the broadening effect is not too serious such that $\beta_w(z)\approx$1 for all $z$, the ground-state coherence $\sigma^{(w)}_{sg,j}(z,t_w)$ can be well approximated by a Gaussian waveform. This approximation is valid if $T_p\delta\omega_w\gg1$. This condition is satisfied for a high enough optical depth and a strong enough coupling field to satisfy $\eta\equiv T_d/T_p\approx$2.5 for storing the major part of the probe pulse inside the medium\cite{YF18}. Under such conditions, $\beta_w(z)$ is approximated by its middle value with $z=v_wt_w$. Thus, $\sigma_{sg,j}^{(w)}(z,t_w)$ can be approximated by
\begin{equation}
\begin{aligned}
&\sigma_{s,j}^{(w)}(z)\equiv\sigma^{(w)}_{sg,j}(z,t_w)\approx\\
&\frac{-\mathcal{E}_0p_jg_pR_j^p}{\Omega_w}\frac{1}{\beta_w(L_w)}exp\left[-\frac{2ln2(z-L_w)^2}{(L_w\beta_w(L_w))^2}\right],
\label{s_sz1}
\end{aligned}
\end{equation}
where $L_w=v_wT_p$, which specifies the spatial length of the probe pulse in the medium. To store nearly all the probe pulse in the medium, $L_w$ must be shorter than the medium length $L$, i.e. $L_w<L$. Under such a condition, the $\sigma_{s,j}(z)^{(w)}$ can be considered as have a nearly complete Gaussian distribution in the atomic medium. For later use, we define a parameter $\kappa=t_w/T_p$. Eq.(\ref{s_sz1}) will be used as the initial condition in the next subsection to calculate the converted field in the reading process. 

\subsection{\label{sec2}Reading process}
After the writing process at time $t=t_w$, we consider the reading process at time $t=t_w+t_s=t_r$ when the reading field is turned on to convert the stored coherence into the converted field. For simplicity, we do not consider the decay of the stored coherence during the storage time $t_s$. The initial condition of the ground state coherence is $\sigma^{(w)}_{sg,j}(z,t'=0)=\sigma^{(w)}_{s,j}(z)$ for each EIT-subsystem, where $t'=t-t_r$. The optical Bloch equations for the reading process under the ideal case of $\gamma_{sg,j}= 0$ are:
\begin{equation}
\begin{aligned}
&\frac{\partial}{\partial t'}\sigma_{e'g,j}=\frac{i}{2}a_{r,j}\Omega_r\sigma_{sg,j}^{(r)}+\frac{i}{2}a_{c,j}p_{j}g_c\mathcal{E}_c-\gamma_{e'g,j}\sigma_{e'g,j},
\label{si41eq}
\end{aligned}
\end{equation}
\begin{equation}
\frac{\partial}{\partial t'}\sigma_{sg,j}^{(r)}=\frac{i}{2}a_{r,j}\Omega_r^*\sigma_{e'g,j},
\label{si21eq}
\end{equation}
where the index $j=1,2\dots,M-1,M$ denotes each EIT-subsystem. $\gamma_{e'g,j}=\Gamma_r/2$ if spontaneous decay is the dominant decoherence mechanism, $g_c=\mu_{eg}\sqrt{\omega_c/2\hbar\epsilon_0V}=\sqrt{\alpha_c\Gamma_rc/2LN}$ is the coupling constant for the field $\mathcal{E}_c$ with optical depth $\alpha_c$ and the dipole moment $\mu_{eg}$, $a_{c,j}$ and $a_{r,j}$ are the Clebsch-Gordan coefficients for the converted and reading transition, and $p_j$ is the atomic population in the $j$-th ground state $|g\rangle$. The Maxwell equation for the converted light field is: 
\begin{equation}
\left(\frac{\partial}{\partial z}+\frac{1}{c}\frac{\partial}{\partial t'}\right)\mathcal{E}_c=\frac{ig_cN}{c}\sum_ja_{c,j}\sigma_{e'g,j},
\label{MSE}
\end{equation}
In the reading process, the converted light is reconstructed from the reading field $\Omega_r$ and the initially stored ground-state coherence $\sigma^{(w)}_{s,j}(z)$. For analytical simplicity, we consider a sudden turn-on of the reading field at time $t'=0^+$ and ignore the turn-on process. In reality, there is a finite turn-on time for the writing field which is considered in the numerical calculation. As discussed in Ref.\cite{PhysRevA.64.043809}, there is almost no difference on the results between these two situations. Using Eqs.(\ref{si41eq}) and (\ref{si21eq}), the evolution of the ground-state coherence $\sigma_{sg,j}^{(r)}(z,t)$ is as follows,
\begin{equation}
\partial^2_{t'}\sigma_{sg,j}^{(r)}+\gamma_{e'g,j}\partial_{t'}\sigma_{sg,j}^{(r)}+\frac{|a_{r,j}\Omega_r|^2}{4}\sigma_{sg,j}^{(r)}=-\frac{a_{r,j}\Omega_r^*a_{c,j}p_jg_c\mathcal{E}_c}{4}.
\label{EOMsi}
\end{equation}
The solution of $\sigma_{sg,j}^{(r)}(z,t')$ in Eq.(\ref{EOMsi}) is given by a general solution plus a particular solution,
\begin{equation}
\begin{aligned}
\sigma_{sg,j}^{(r)}(z,t')=\chi_{j}(t')\sigma^{(w)}_{s,j}(z)+\sigma_{sg,j}^{{p}}(z,t'),
\end{aligned}
\label{si21sol}
\end{equation}

\begin{equation}
\chi_{j}(t')=\left[\frac{\gamma_{e'g,j}}{2\delta_j} sin(\delta_{j} t')+cos(\delta_{j} t')\right ]e^{-\frac{\gamma_{e'g,j} t'}{2}},
\label{chi}
\end{equation}

\begin{widetext}
\begin{equation}
\begin{aligned}
\sigma_{sg,j}^{{p}}(z,t')
=-\frac{a_{r,j}\Omega_r^*p_ja_{c,j}g_c}{4\delta_j}\int_{0}^{t'}e^{-\frac{\gamma_{e'g.j}(t'-t'')}{2}}sin[\delta_j(t'-t'')]\mathcal{E}(z,t'')dt''\\
=-\frac{\sqrt{2\pi}a_{r,j}\Omega_r^*p_ja_{c,j}g_c}{4\delta_j}&\left[e^{-\frac{\gamma_{e'g,j} t'}{2}}sin(\delta_{j} t')*\mathcal{E}_c(z,t')\right](t'),  
\end{aligned}
\label{psol}
\end{equation}
\end{widetext}
where $\delta_j=\left(|a_{r,j}\Omega_r|^2-\gamma_{e'g,j}^2\right)^{1/2}/2$ and the notation $\left[f(t')*g(t')\right](t')=\frac{1}{\sqrt{2\pi}}\int_{0}^{t'}f(t'-t'')g(t'')dt''$ is the convolution of the two functions $f(t')$ and $g(t')$ in a finite range from $0$ to $t'$.
The coefficient $\chi_{j}(t')$ represents the oscillation of the ground-state coherence with time due to a constant driving of the reading field. The particular solution represents the contribution on the ground-state coherence due to the generated conversion field. 

By inserting Eqs.(\ref{si21sol}),(\ref{chi}), and (\ref{psol}) into Eq.(\ref{si21eq}), one obtains the expression for $\sigma_{e'g,j}$. By inserting this relation into the Maxwell equation (Eq.\ref{MSE}) and then Fourier transforming it to $\omega$-space and applying the convolution theorem, the $\omega$-space Maxwell equation becomes,  
\begin{equation}
\begin{aligned}
\left(\frac{\partial}{\partial z}+f^{r}(\omega)\right)\mathcal{E}_c(z,\omega)&=\frac{1}{\sqrt{2\pi}}\frac{2g_cN}{c\Omega_r^*}\sum_{j}R_j^{c}A^{r}_j(\omega)\sigma^{(w)}_{s,j}(z).
\end{aligned}
\label{MSE3}
\end{equation}
The coefficient in Eq.(\ref{MSE3}) are given by
\begin{equation}
\begin{aligned}
A^r_j(\omega)=-\left[1-\frac{4\omega^2+2i\Gamma_r\omega}{|a_{r,j}\Omega_r|^2}\right]^{-1},
\end{aligned}
\end{equation}
\begin{equation}
\begin{aligned}
f^r(\omega)=-\frac{i\omega}{c}\left[1-\frac{2g_c^2N}{|\Omega_r|^2}\sum_jp_j(R_j^c)^2A_j^r(\omega)\right],
\end{aligned}
\end{equation}

where $R^c_{j}=a_{c,j}/a_{r,j}$ is the ratio of the Clebsch-Gordon coefficients for the converted and reading transition. Eq.(\ref{MSE3}) describes how the conversion field is generated from the ground-state coherences and how it evolves during propagation in the medium. With the initial condition of $\mathcal{E}_c(z=0,\omega)=0$, the solution of Eq.(\ref{MSE3}) for the converted field in $\omega$-space is,
\begin{equation}
\begin{aligned}
\mathcal{E}_c(z,\omega)&=\frac{2g_cN}{c\Omega_r^*}\sum_jR_j^cA^r_j(\omega)\frac{1}{\sqrt{2\pi}}\int_0^z\sigma^{(w)}_{s,j}(z')e^{-f^r(\omega)(z-z')}dz'\\& =\frac{2g_cN}{c\Omega_r^*}\sum_jR_j^cA^r_j(\omega)\left[exp\left(-f^r(\omega)z\right)*\sigma^{(w)}_{s,j}(z)\right](z).
\label{Ec}
\end{aligned}
\end{equation}
For the given initial ground-state coherence $\sigma^{(w)}_{s,j}(z)$ determined by the writing process (Eq. (\ref{s_sz1})), the converted field can be calculated by Eq.(\ref{Ec}). By performing the inverse Fourier transform on Eq.(\ref{Ec}), one can obtain the time-domain waveform of the converted field.

\subsection{\label{sec3}Characteristic of the converted field}
\begin{figure*}[th]
\centering
\includegraphics[width=1\textwidth]{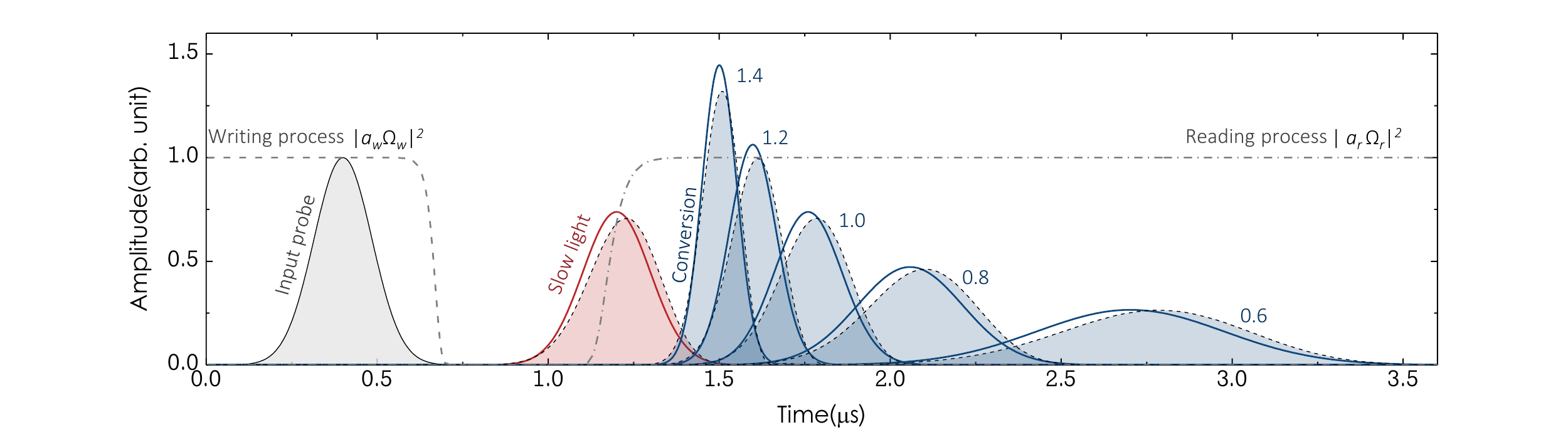}
\caption{Demonstration of the conversion process in the pulse region. Here we consider the conversion process with a single Zeeman state. $\Gamma_r=\Gamma_w=2\pi\times4.56$MHz and $a_p=a_c$. With the optical depth of $a^2_p\alpha_p=500$ and the factor $\eta$ is 4 to satisfy the condition for storing almost all the light pulse. The gray area denotes the input probe pulse with $T_p=0.2\mu s$. The red (blue) line and area indicate the analytic solution of Eq.(\ref{Ec_w}) and the numerical simulation for the slow light (conversion) process. We show various converted pulses which correspond to different reading field intensities. The numbers shown beside the converted pulses denote the ratio of $|a_r\Omega_r|/|a_w\Omega_w|$. 
}
\label{pulse}
\end{figure*}

To find an approximate analytic form for the converted field, we substitute the result for Eq.(\ref{s_sz1}) into Eq.(\ref{Ec}). Considering $f^r(\omega)$ to the second-order dispersion term of $\omega$ and assuming the adiabatic condition (i.e. $A_j^r(\omega)\approx -1$, the solution of the converted field in the frequency domain and out of the medium is,
\begin{widetext}
\begin{equation}
\begin{aligned}
\mathcal{E}_c(L,\omega)\cong\frac{g_cg_pN\mathcal{E}_0}{\sqrt{ln2}c\Omega_r^*\Omega_w}\sum_jR_j^cR_j^pp_jL_h^wexp\left[\frac{i\omega}{v_r}(L-L_w)-\frac{[L_h^w\beta(L_w)]^2\beta^2_r(L)}{8ln2v_r^2}\omega^2\right].
\label{Ec_w}
\end{aligned}
\end{equation}
\end{widetext}
The factor $\beta_r$ is given by
\begin{equation}
\begin{aligned}
\beta_r(L)&=\left[1+(\frac{4ln2}{\delta\omega_r\beta_w(L_w)T_p})^2\frac{v_r^2(L-L_w)}{v_w^2L}\right]^{1/2},\\
\end{aligned}
\label{betar}
\end{equation}
where $v_r$ is the group velocity of the converted field in the medium and $\delta\omega_r$ is the EIT-bandwidth of the converted transition, which correspond to $v_w$ and $\delta\omega_w$, respectively, by replacing the subscript $w\rightarrow r$ and $p\rightarrow c$ in Eqs. (\ref{vw}) and (\ref{deltaw}). We assume that the optical depth is larger enough (e.g. $>\sim $100) and the parameters $\kappa=t_w/T_p$ and $\eta=T_d/T_p$ are suitably chosen such that the major part of the probe pulse is stored inside the medium\cite{YF18} during the writing process. These conditions allow us to neglect the higher order terms of $O(\omega^3)$ and above in Eq. (\ref{Ec_w}) and to approximate an integral to achieve Eq. (\ref{Ec_w}) by using $\int_{-\infty}^{\infty}e^{-au^2}du=\sqrt{\pi/a}$. After applying the inverse Fourier transform on Eq.(\ref{Ec_w}), the approximate analytic formula of the converted field in the time domain can be obtained. An example demonstrating the conversion process, with the approximate analytic formula and the numerical calculation based on the Maxwell-Bloch equations is shown in Fig.\ref{pulse}. Comparison shows that Eq.(\ref{Ec_w}) closely approximates well with the numerical simulation, with a slight deviation due to neglecting the higher order terms.  

We further explore some properties of the converted field based on Eq.(\ref{Ec_w}). It is useful to know the spectral distribution of  the converted light because it tells us how to manipulate its spectral properties by memory-based conversion. The spectral distribution is also related to the quantum fidelity in the frequency domain\cite{PhysRevA.88.013823}. The FWHM bandwidth $\Delta\omega_c(L)$ of the spectral power density $S_r(z=L,\omega)=|\mathcal{E}_c(L,\omega)|^2$ of the converted field out of the medium is, 

\begin{equation}
\begin{aligned}
&\Delta\omega_c(L)=\frac{1}{\beta_r(z)\beta_w(L_w)}\left|\frac{\Omega_r}{\Omega_w}\right|^2\frac{\sum_jg_p^2p_j(R_j^p)^2}{\sum_jg_c^2p_j(R_j^c)^2}\Delta\omega_0.
\label{Dw_c}
\end{aligned}
\end{equation}
From Eq. (\ref{Dw_c}), it is evident that the bandwidth of the converted field is determined by the intensity ratio of the writing field to the reading field and the atomic parameters of the relevant transitions. By adjusting the reading field intensity, the spectral bandwidth or the temporal waveform of the converted light can be easily manipulated. 

Another important quantity is the conversion efficiency defined by taking the energy ratio between the converted light to that of the input light. By integrating all the frequency composition of $S_r(z=L,\omega)$ and normalizing to that of the input probe field, we obtain the conversion efficiency as follows:
\begin{equation}
\begin{aligned}
&\xi^T_c=\left[\frac{1}{\beta_w(L_w)\beta_r(L)}\right]\left[\frac{\left|\sum_jp_jR_j^pR_j^c\right|^2}{\sum_jp_j(R_j^p)^2\sum_jp_j(R_j^c)^2}\right]\equiv \xi_1\xi_2.
\label{eff_c}
\end{aligned}
\end{equation}
The two terms shown in the brackets in Eq.(\ref{eff_c}) are denoted as $\xi_1$ and $\xi_2$, respectively. The first term ($\xi_1$) can be understood as the finite EIT-bandwidth effect\cite{YF18}. From the relations of $\beta_w$ and $\beta_r$ (Eqs. (\ref{betaw}) and {\ref{betar}}), it is evident that for larger EIT transparent bandwidths these two factors approach unity. More accurately, these two parameters are related to the time-bandwidth product of the probe and conversion transition ($T_p\delta\omega_{w}$). In the cases to compress the major part of the probe pulse into the medium, the group delay $T_d=\eta T_p$ with $\eta$ being a constant of $\sim 2.3-3$, depending on the optical depth\cite{YF18}. Therefore, the two parameters $\beta_w$ and $\beta_r$ are related to the delay-bandwidth product of the probe and conversion transition, which is dependent on the optical depth of each transition\cite{YF18}.  

The second term $\xi_2$ is related to the ground-state coherence mismatch between the writing and reading phase and has been discussed in Refs.\cite{OptCommun.209.149,OptLett.31.3511,PhysRevA.69.043801,PhysRevA.75.013812}. According to the Cauchy-Schwarz inequality, one has $\left[\sum_j(\sqrt{p_j}R_j^p)(\sqrt{p_j}R_j^c)\right]^2\leq \left[\sum_j(\sqrt{p_j}R_j^p)^2\right]\left[\sum_j(\sqrt{p_j}R_j^c)^2\right]$. Therefore, $\xi_2$ is always smaller than or equal to unity. The equality holds when $R_j^p/R_j^c$ is a constant for each EIT subsystem or all the population occupy a single Zeeman state. The reason for this term and under what conditions the equality holds have been well explained in Ref.\cite{PhysRevA.75.013812}. Here, we briefly mention the essential point. From Eq. (\ref{sig_s}), it is evident that the ground-state coherence of the $j^{th}$ EIT subsystem is $\sigma_{eg,j}^{(w)}=-p_jR_j^p\frac{g_p\mathcal{E}_p(\omega=0)}{\Omega_w}$ in the adiabatic limit (i.e. $A_j^w(\omega)\simeq -1$). A similar relation holds for the reading phase in the adiabatic limit, i.e. 
\begin{equation}
\begin{aligned}
\sigma_{sg,j}^{(r)}=-p_jR_j^c\frac{g_c\mathcal{E}_c(\omega=0)}{\Omega_r}.
\end{aligned}
\label{gscoh}
\end{equation}
If the ratio of $R_j^c/R_j^p$ is different for each EIT subsystem, the ground-state coherence for the reading phase at the initial time of retrieval cannot simultaneously satisfy Eq. (\ref{gscoh}) for all subsystem with one given converted field $\mathcal{E}_c$. Therefore, some of the $\sigma_{sg,j}^{(r)}$ may change in order to reach a condition where Eq. (\ref{gscoh}) is valid again for all subsystems for a given converted field in the adiabatic limit. According to Eq. (\ref{si21eq}), the variation of Re[$\sigma_{sg,j}^{(r)}$] must be accompanied by nonzero Im[$\sigma_{e'g,j}$], which lead to energy loss of the converted light during this process. If $R_j^c/R_j^p$ is a constant for all subsystems, there exists one converted field such that Eq. (\ref{gscoh}) is satisfied for all subsystems and all $\sigma_{eg,j}^{(r)}$ remain the same during reading. This condition also holds if the entire population is prepared in a single Zeeman state. Thus, there is no energy loss for the converted field under such conditions\cite{PhysRevA.75.013812}. Such an example exists for wavelength conversion between the $D_1$ and $D_2$ line of alkali atoms with all the laser fields having the same $\sigma^{+}$ (or $\sigma^{-}$) polarization\cite{PhysRevA.75.013812}.      

Under the condition that $\xi_2$ equals unity, there is still a loss due to the finite EIT-bandwidth effect characterized by the $\xi_1$ factor. To further explore the bandwidth effect, in the next section, our discussion focuses on the case where the entire population is in a single Zeeman state, so the conversion efficiency is only affected by $\xi_1$.

\section{\label{sec4}Population in a single Zeeman state}
For the case in which all the population is in a single Zeeman state, the conversion efficiency is determined by $\xi_1$ alone. The efficiency $\xi_1$ depends on two factors $\beta_w(L_w)$ and $\beta_r(L)$ during the writing and reading process, respectively. With the definition $\kappa=t_w/T_p$ and $\eta=T_d/T_p$ and using Eqs.(\ref{vw})(\ref{deltaw})(\ref{betaw})(\ref{Td})(\ref{betar}) and the corresponding relations for $v_r$ and $\delta\omega_r$, we obtain the following two relations,
\begin{equation}
\beta_w(L_w)=\left[1+16ln2\frac{\eta\kappa}{D_p}\right]^{1/2},  
\label{betawsim}
\end{equation}
\begin{equation}
\beta_r(L)=\left[1+16ln2\frac{\eta(\eta-\kappa)}{D_c}\right]^{1/2}, 
\label{betarsim}
\end{equation}
where $D_p=a_p^2\alpha_p$ and $D_c=a_c^2\alpha_c$ are the optical depth of the medium for the probe and conversion transition, respectively. The conversion efficiency is given by $\xi_c^T=\xi_1=1/(\beta_w(L_w)\beta_r(L))$, which is determined by the parameters $\eta$, $\kappa$, $D_p$ and $D_c$. Note that the conversion efficiency does not depend on the Rabi frequencies of the writing and reading field, $\Omega_w$ and $\Omega_r$. We remind the reader that these two approximate relations are valid under the assumption that the optical depths are large enough (e.g. $>~100$), $\kappa>\sim1.1$ and $\eta>\sim2.5$ such that the major part of the probe pulse can be stored in the medium\cite{YF18}.     

Because we consider the writing and reading process in two different EIT channels, it is interesting to know what is the net difference in the conversion efficiency due to the different atomic properties of the two EIT channels in the reading process. To quantify this comparison, we introduce a parameter known as the relative conversion efficiency $\xi^R_c$ defined by $\xi^R_c=\xi^T_c/\xi^T_w$, where $\xi^T_w$ is the storage efficiency of the probe field written and read in the original EIT channel. The efficiency due to the writing process has been normalized away in $\xi^R_c$. $\xi^R_c$ is the efficiency ratio of the reading process of the conversion EIT channel to that of the original EIT channel, which reads as follows,     
\begin{equation}
\begin{aligned}
&\xi^R_c=\left[\frac{1+\frac{16ln2(1-\kappa/\eta)\eta^2}{a_p^2\alpha_p\beta^2_w(L_w)}}{1+\frac{16ln2(1-\kappa/\eta)\eta^2}{a_c^2\alpha_c\beta^2_w(L_w)}}\right]^{1/2}.
\label{xi_c}
\end{aligned}
\end{equation}

\begin{figure}[t]
\centering
\includegraphics[width=0.52\textwidth]{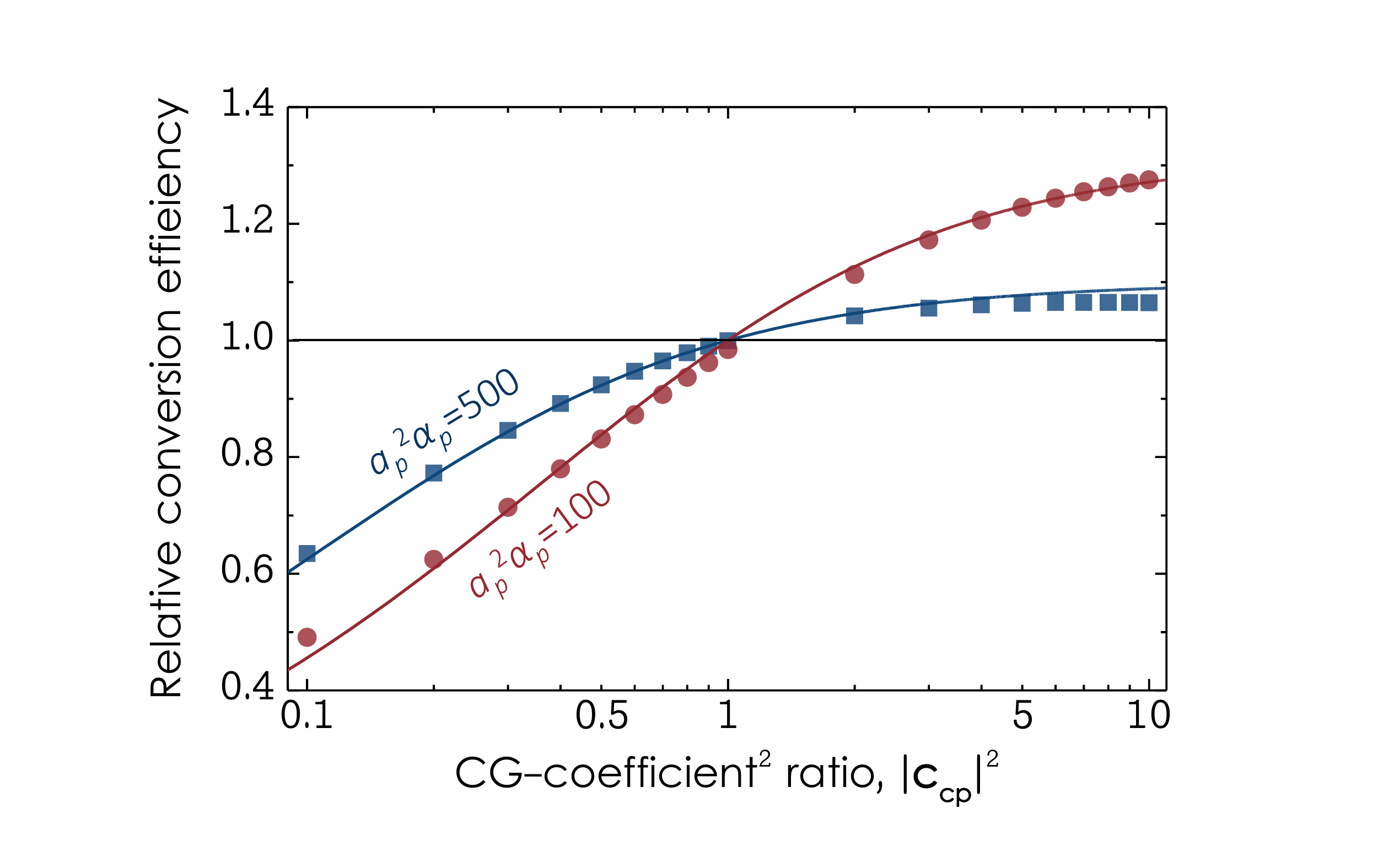}
\caption{The relative conversion efficiency versus the CG-coefficient ratio $|c_{cp}|^2$. The red dots and blue squares denote the results of the numerical simulation with optical depths $a_p^2\alpha_p$ in the writing channel equal to 100 and 500, respectively. The solid lines represent the Eq.(\ref{eff_c}) results. The parameters values are $\eta=4$ and $\kappa=1.35$.}
\label{eff_num}
\end{figure}

For a quantitative discussion, consider a practical example of the memory-based conversion system as shown in Fig. \ref{process2} for the cesium $D_1$-line in which the probe and writing fields drive the $\sigma^+$($\sigma^-$) transitions and the conversion and reading fields drive the $\sigma^-$($\sigma^+$) transitions. Assume that the entire population is prepared in a single Zeeman state. With these settings, we have $\Gamma_r=\Gamma_w$ and $\alpha_c=\alpha_p$. The relative efficiency can be obtained by inserting these relations into Eq. (\ref{xi_c}). We define a parameter $c_{cp}=a_c/a_p$, which is the ratio of the Clebsch-Gordon coefficient of the conversion transition to that of the probe transition. Its square $|c_{cp}|^2$ is the optical depth ratio of the conversion transition to that of the probe transition. Fig. \ref{eff_num} depicts $\xi_c^R$ versus $|c_{cp}|^2$ for two different values of optical depth. For the case of $|c_{cp}|^2>1$, $\xi_c^R>1$ means that the energy of the converted light field is greater than that of the retrieved probe pulse in the original EIT channel. For the case of $|c_{cp}|^2<1$, the situation is the opposite. The solid lines in Fig.\ref{eff_num} are based on the approximate formula of Eq. (\ref{xi_c}) and the data points are based on the full numerical simulation with the Maxwell-Bloch equations. The parameters used in Fig. \ref{eff_num} are $\eta=4$ and $\kappa=1.35$. It can be seen that the analytic formula matches the numerical calculation well given these parameters. 

\begin{figure}[t]
\centering
\includegraphics[width=0.52\textwidth]{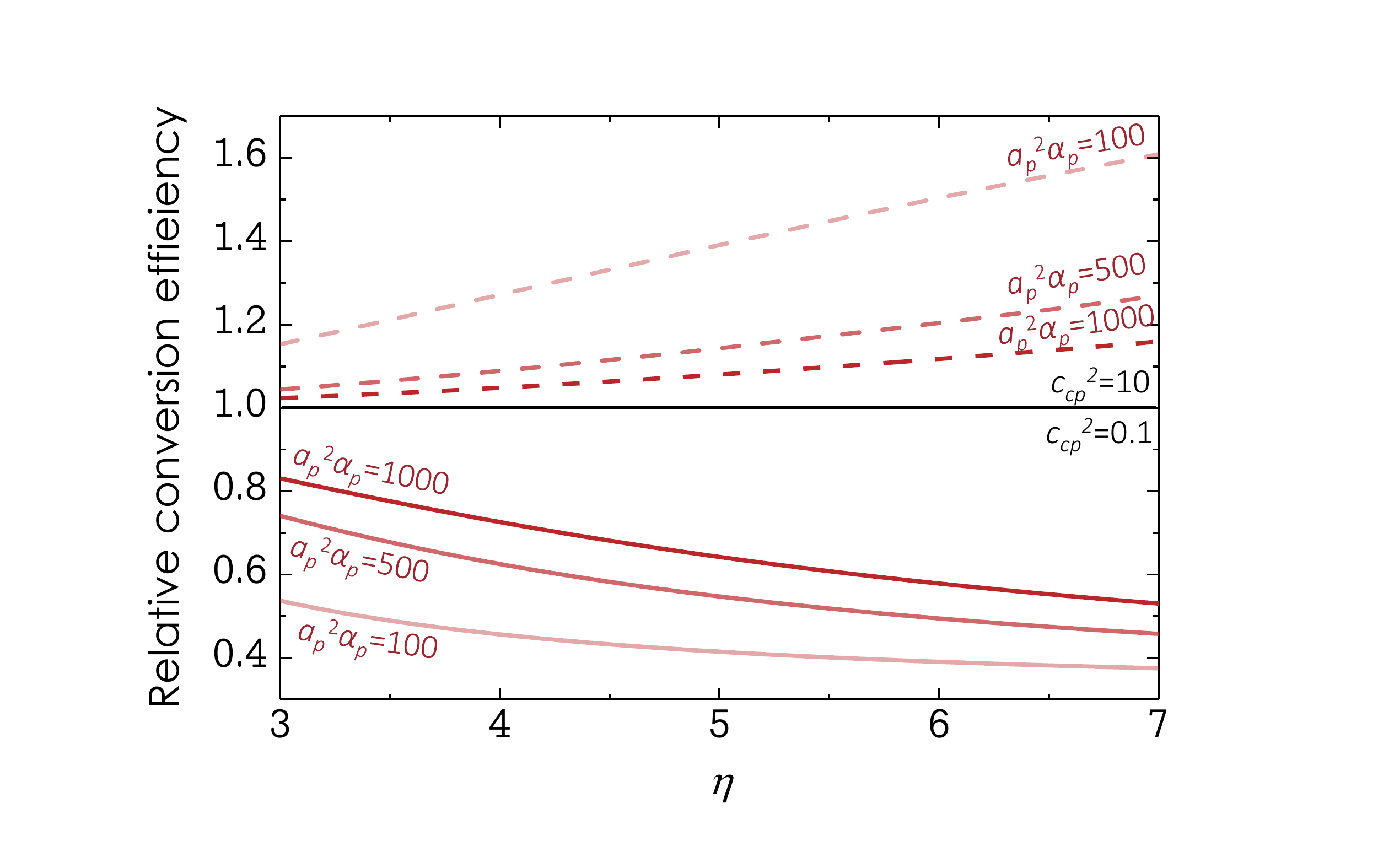}
\caption{The behavior of $\xi_c^R$ versus the parameter $\eta$. The dashed lines denote the behaviors with $|c_{cp}|^2=10$ for various optical depths shown next to the lines. The solid lines denote the behaviors with $|c_{cp}|^2=0.1$. When $|c_{cp}|^2>1$,  $\xi_c^R$ increases as $\eta$ increases and  when $|c_{cp}|^2<1$, $\xi_c^R$ decreases as $\eta$ increases.
}
\label{eta}
\end{figure}

\begin{figure*}[t]
\centering
\includegraphics[width=1\textwidth]{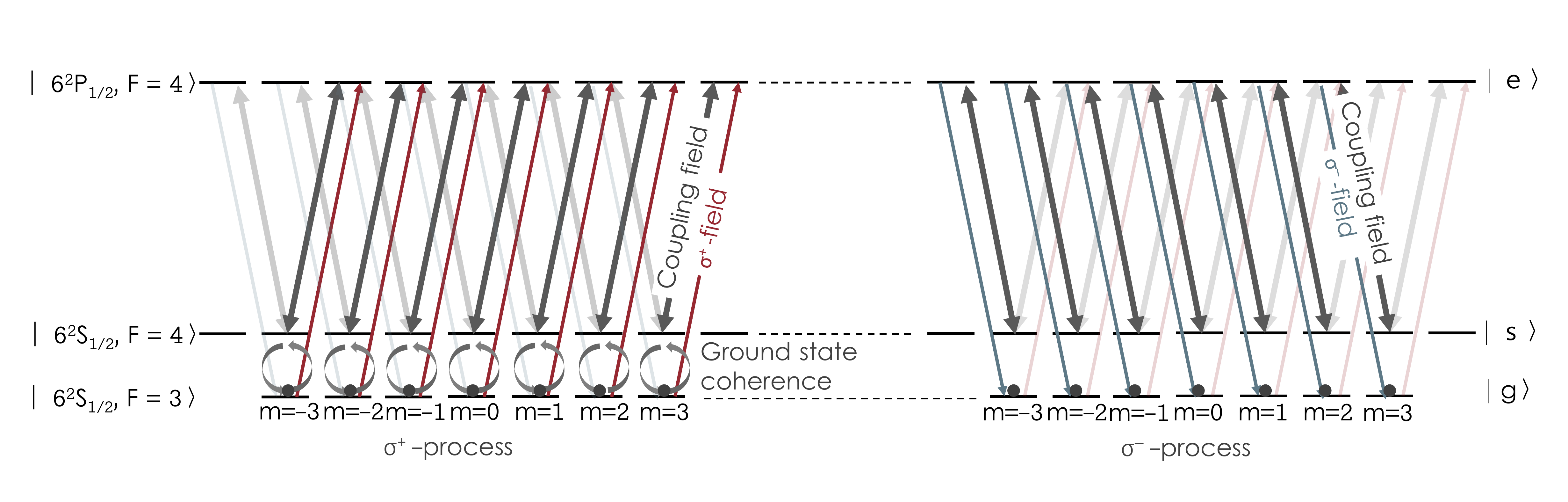}
\caption{Energy level diagram of the memory-based optical converter based on the cesium $D_1$-line transitions. There are fourteen Zeeman ground states and nine excited Zeeman states involved. With different polarization of the input light field, this converter can realize  conversion of polarization from $\sigma^+$ to $\sigma^-$ or from $\sigma^-$ to $\sigma^+$.}
\label{process2}
\end{figure*}

For a specific atomic transition chosen for a conversion system, the CG-coefficient ratio $|c_{cp}|^2$ is basically fixed. To vary $|c_{cp}|^2$, one has to choose a different atomic transition which involves lasers at different wavelengths or prepare the population in different single Zeeman state. Both tasks are nontrivial. It is therefore difficult to do an intensive experimental test on the $|c_{cp}|^2$ dependence of $\xi_c^R$. To test the relation of $\xi_c^R$, there is one more parameter $\eta$ that one can vary by changing the intensity of the writing field during the writing process. Fig.\ref{eta} depicts the behavior of $\xi_c^R$ versus $\eta$ for different $|c_{cp}|^2$ and optical depths. It can be seen that for $|c_{cp}|^2 >1$, $\xi_c^R(\eta)$ are all larger than unity and $\xi^R_c$ is larger for a larger $\eta$. For $|c_{cp}|^2 <1$, $\xi_c^R(\eta)$ are all less than unity and $\xi^R_c$ is smaller for a larger $\eta$. $\xi^R_c$ approaches unity for a smaller $\eta$ in all cases. This is understandable and is explained below. A smaller $\eta$ is accompanied by a stronger intensity and thus a wider EIT bandwidth. This means that the ratio of the pulse spectral bandwidth to the EIT bandwidth is decreasing and thus the finite EIT bandwidth effect becomes less important. In other words, the situation approaches the continuous wave case and the retrieval efficiency in the converted and the original EIT channel approaches unity, as well as their ratio $\xi^R_c$.

In a real experiment, it may not be easy to prepare all of the population in a single Zeeman state, especially for an optically dense medium due to the radiation trapping effect\cite{Szymaniec2013}. It is helpful to consider memory-based conversion for the condition wherein the atomic population is distributed among multi-Zeeman states. We discuss such a situation in the next section.

\section{\label{sec5}Dispersed population distribution in multi-Zeeman states}

We then consider the case for a memory-based conversion system with a dispersed population distributed among the various Zeeman states, as shown in Fig. \ref{process2}. For a given Zeeman population distribution, we consider two situations with the writing channel driving the $\sigma^+$ transitions and the reading channel driving the $\sigma^-$ transitions or the opposite, which are denoted as $\sigma^+ \rightarrow \sigma^-$ and $\sigma^- \rightarrow \sigma^+$, respectively. It should be noted that the Zeeman population distribution affects the effective optical depth for both the writing and reading channels. For example, the effective optical depth for the probe transition is $\sum_jp_ja^2_{+,j}\alpha_p$, where $a_{+,j}$ is the Clebsch-Gordon coefficient of the $\sigma^+$ probe transition in each EIT subsystem. To simplify the comparison, we consider the relative conversion efficiency $\xi_c^R$. From Eq. (\ref{eff_c}) and the definition of $\xi^R_c$, we have    
\begin{widetext}
\begin{equation}
\begin{aligned}
\xi^R_c=\frac{\left|\sum_jp_jR_j^pR_j^c\right|^2}{\sum_jp_j(R_j^p)^2\sum_jp_j(R_j^c)^2}
\left[1+\frac{16ln2(\eta-\kappa)}{\beta^2_w(L_w)}\frac{\sum_jp_j(R_j^p)^4/(a_{p,j}^2\alpha_p)}{(\sum_jp_j(R_j^p)^2)^2}\right]^{\frac{1}{2}}
\left[1+\frac{16ln2(\eta-\kappa)}{\beta^2_w(L_w)}\frac{\sum_jp_j(R_j^c)^4/(a_{c,j}^2\alpha_c)}{(\sum_jp_j(R_j^c)^2)^2}\right]^{-\frac{1}{2}}.
\label{xi_i_c2}
\end{aligned}
\end{equation}
\end{widetext}

\begin{table}[b]
    \centering
    \begin{tabular}{lccccccc}
    \hline
    $j$        & -3   & -2   & -1   & 0    & 1   & 2   & 3   \\ \hline
    $R_j^-$    & $\sqrt{7}$    & $\sqrt{3}$   & $\sqrt{5/3}$   & 1   & $\sqrt{3/5}$   & $\sqrt{1/3}$ & $\sqrt{1/7}$\\ \hline
    $R_j^+$   & $-\sqrt{1/7}$    & $-\sqrt{1/3}$   & $-\sqrt{3/5}$   & -1   & $-\sqrt{5/3}$   & $-\sqrt{3}$ & $-\sqrt{7}$\\ \hline
    \end{tabular}
    \caption{The ratio of CG-coefficient of cesium $D_1$-line with the transition of $\sigma^\pm$.}\label{ratio_CG}
\end{table}

\begin{figure}[t]
\centering
\includegraphics[width=0.52\textwidth]{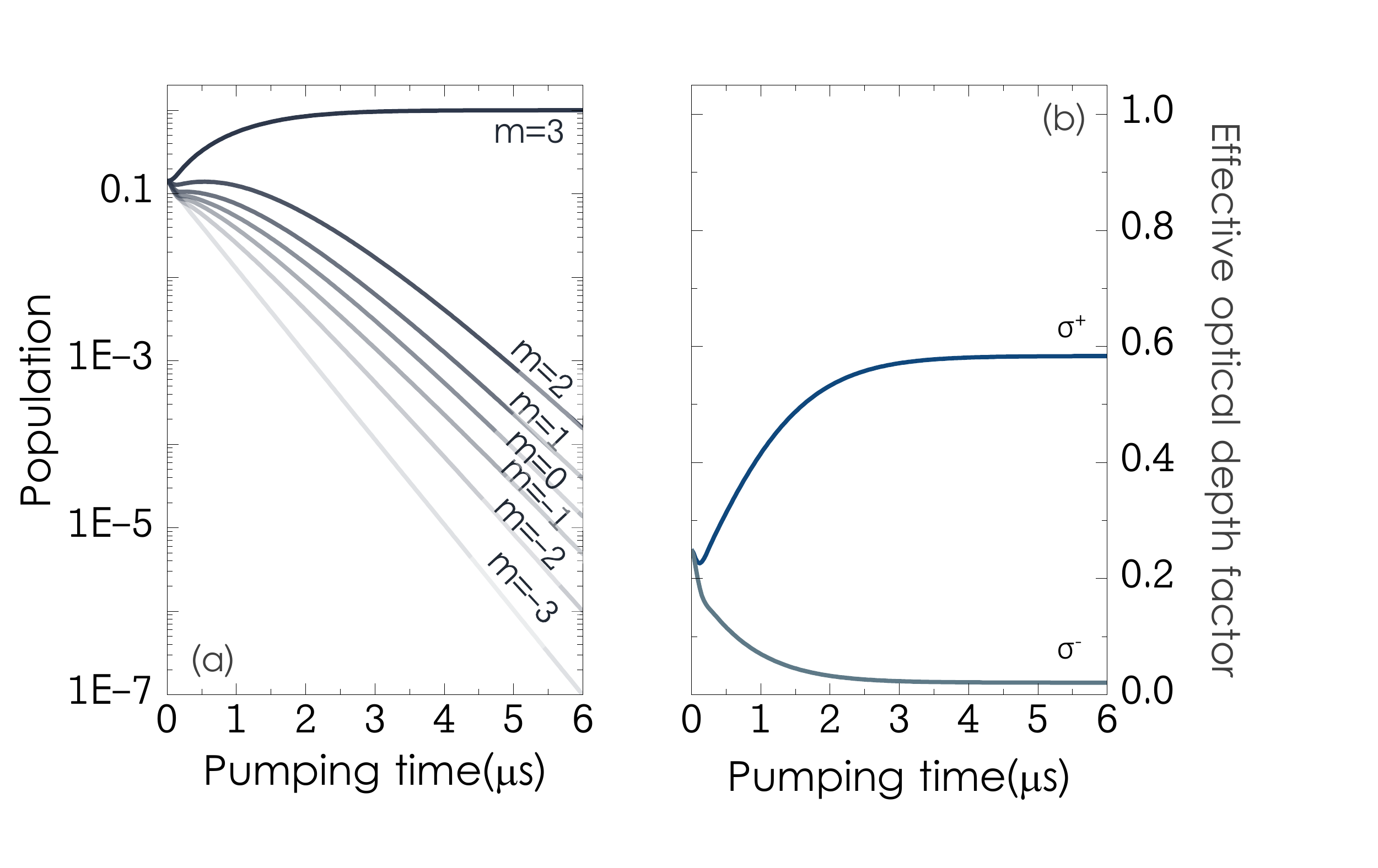}
\caption{The simulation results of optical pumping dynamics. We assume that the initial population is isotropically distributed among all Zeeman states. The optical pumping field is $\sigma^+$-polarized. The population is nearly 100$\%$ in the $|m=3\rangle$  state in the steady state, as shown in (a). (b) The effective optical depth factor for both $\sigma^+$  and $\sigma^-$ for the population distribution in (a).}
\label{pumping}
\end{figure}

To study the variation of $\xi_c^R$ versus different Zeeman population distributions, we perform a numerical simulation of  Zeeman optical pumping and prepare various Zeeman population distributions. The Zeeman optical pumping beam may drive the $\sigma^+$, $\sigma^-$ and/or $\pi$ transition, depending on its polarization (see Appendix). Assume an initial condition where all of the population is isotropically distributed among all Zeeman states of the $6 S_{1/2}, F=3$ ground state. As an example, we first consider a pumping field driving the $\sigma^+$ transition which gradually pumps the whole population towards the $\left|F=3,m=3\right\rangle$ state. The dynamics of the population are shown in Fig.\ref{pumping}. The effective optical depth factor, defined as $\sum p_ja^2_{\pm,j}$, for the transition of $\sigma^+$ and $\sigma^-$ has a different trend, as shown in Fig.\ref{pumping}(b). For a more concentrated population distribution to the $|m=3\rangle$ state, the effective optical depth becomes higher for the $\sigma^+$ transition but lower for the $\sigma^-$ transition. By putting the Zeeman population distribution into 
Eq.(\ref{xi_i_c2}), we can study its dependence on $\xi^R_c$. For reference, we show the ratio of the CG-coefficient for the cesium $D_1$-line in Table.\ref{ratio_CG}\cite{Steck98cesiumd}. 

\begin{figure}[th]
\centering
\includegraphics[width=0.52\textwidth]{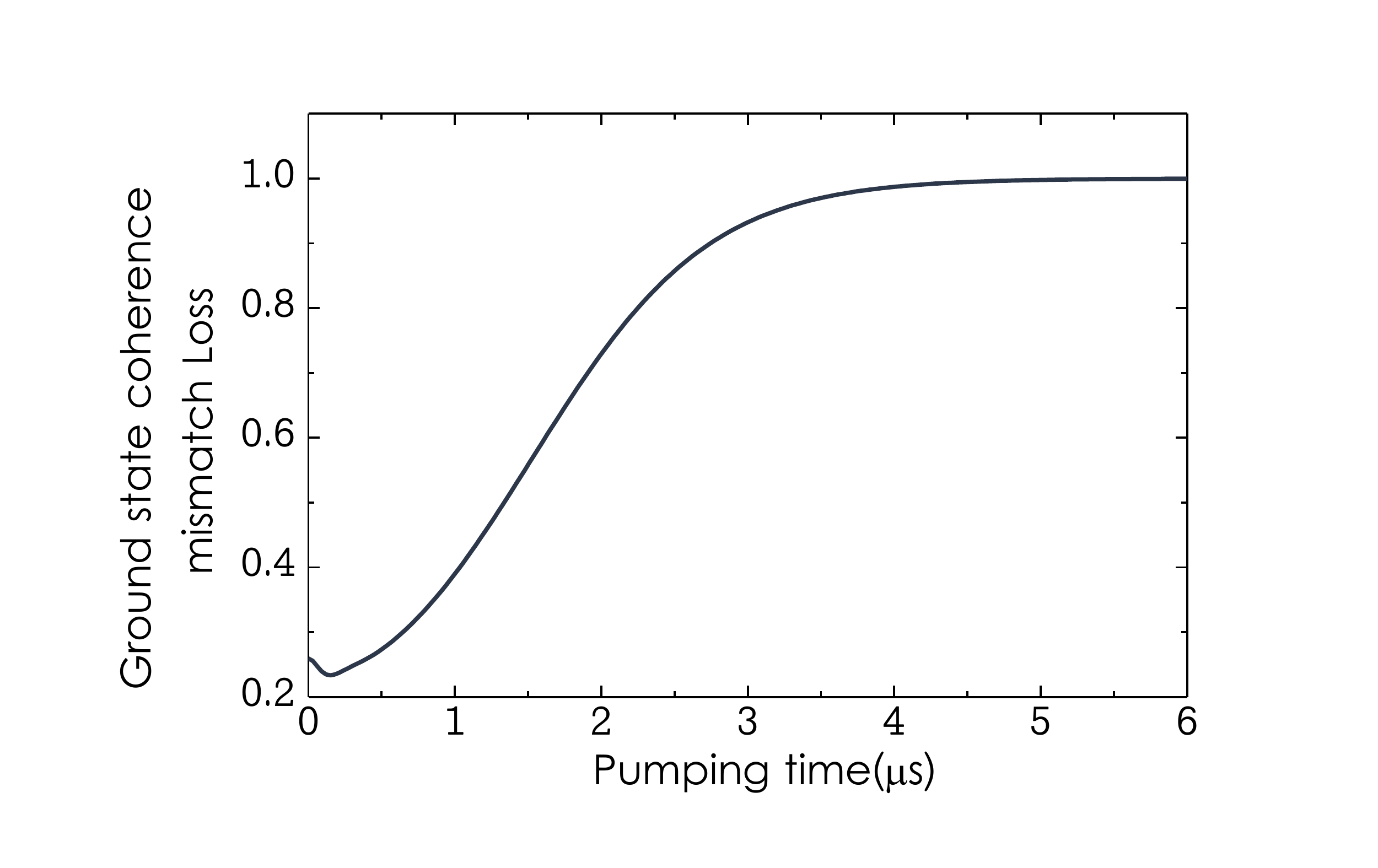}
\caption{The loss factor due to the ground-state coherence mismatch ($\xi_2$) versus the population dynamics. We consider the Table.\ref{ratio_CG} and the simulation results in Fig.\ref{pumping} when estimating the value of the loss. From the isotropic distribution of the population to the highly concentrated distribution, the loss factor follows the pumping time from 26\% almost to $100\%$.}
\label{loss}
\end{figure}

Before demonstrating the behavior of $\xi_c^R$ versus different population distributions, first consider the continuous (CW) probe case, which is free from the finite-bandwidth effect. In the CW limit, the relative conversion efficiency is dominated by the ground-state coherence mismatch factor ($\xi_2$). Along with the pumping time, $\xi_2$ evolves from 0.26 to almost $1$, as shown in Fig.\ref{loss}. This shows that the distribution of the Zeeman population has a serious effect on the conversion efficiency. It is noted that this factor is determined by the atomic parameters only and is independent of whether the conversion process is from $\sigma^+$ to $\sigma^-$ or the opposite. 

\begin{figure}[t]
\centering
\includegraphics[width=0.52\textwidth]{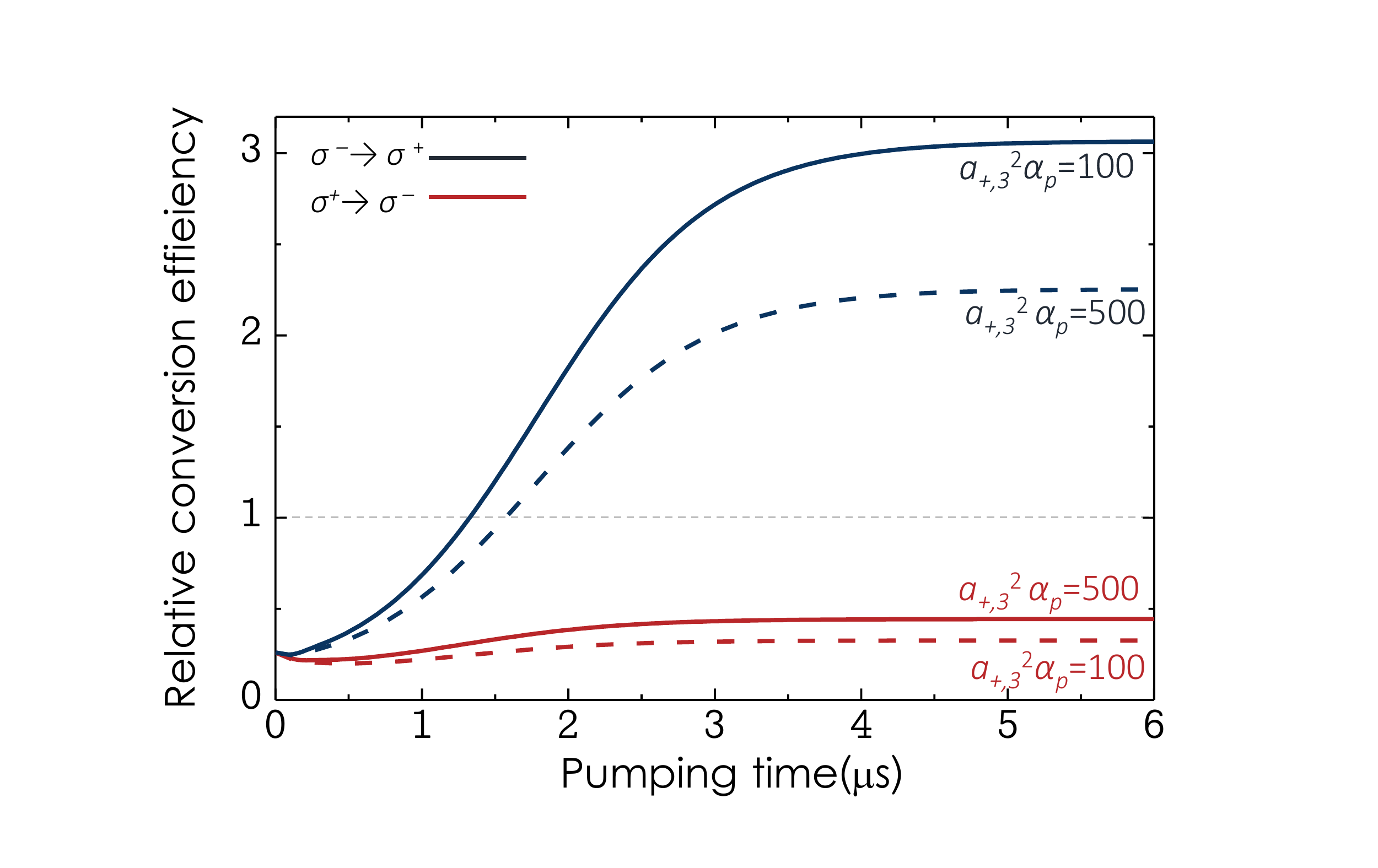}
\caption{The relative conversion efficiency with different optical depths for both the $\sigma^- \rightarrow \sigma^+$ and $\sigma^+ \rightarrow \sigma^-$. As with the pumping time, $\xi_c^I$ varies with the population distribution dynamically. With different conversion type, $\xi_c^I$ presents totally different behavior. In this case, we define the condition of $\eta=4$ and $\kappa=1.35$. }
\label{xi_i_pump}
\end{figure}
\begin{figure}[t]
\centering
\includegraphics[width=0.52\textwidth]{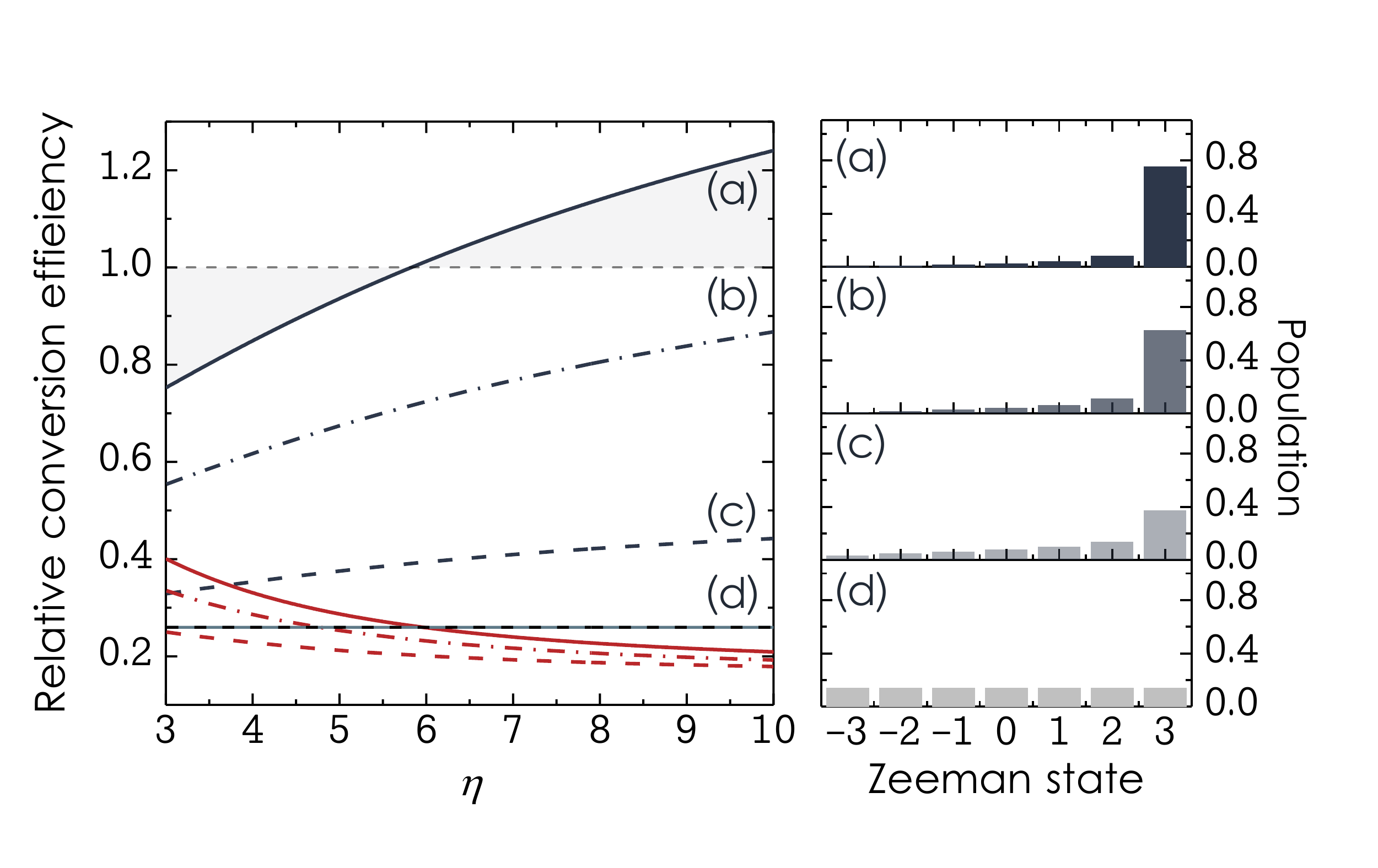}
\caption{The behavior of $\xi_c^R$ versus the factor $\eta$ for different population distribution. Here we set the optical depth $a_{+,3}^2\alpha_p=500$ and $\kappa=1.35$. The dark blue (red) line represents the conversion process of $\sigma^-\rightarrow\sigma^+$ ($\sigma^+\rightarrow\sigma^-$). We consider four different population distributions with a pumping time of 1.6$\mu$s for case (a), 1.2$\mu$s for (b), 0.6$\mu$s for (c) and 0$\mu s$ for (d). The corresponding $\xi_c^R$ versus $\eta$ for the four cases are plotted on the left plot. In case (d), $\xi_c^R$ is a constant with the same values in both conversion types, so the lines overlap. The horizontal line with a value of 1 is used as a reference.}
\label{xi_i_eta}
\end{figure}

Next, we consider the pulse case in which $\xi^R_c$ is affected by both the finite-bandwidth factor ($\xi_1$) and the coherence mismatch factor ($\xi_2$), as shown in Fig.\ref{xi_i_pump}. For $\sigma^- \rightarrow \sigma^+$ conversion, the effective optical depth increases as the population concentrates in the $|m=3\rangle$ state along with the optical pumping, as shown in Fig.\ref{pumping} (b). At an early pumping time, $\xi_c^R$ is dominated by $\xi_2$, which is well below unity, such that $\xi_c^R$ is less than unity although $\xi_1$ could be slightly larger than unity. As the population is pumped towards concentrating in the $|m=3\rangle$ state, $\xi_2$ approaches unity and $\xi_1$ also becomes much larger than unity such that the overall $\xi_c^R$ is larger than unity, as shown in Fig. \ref{xi_i_pump}. In $\sigma^+ \rightarrow \sigma^-$ conversion, the effective optical depth decreases as the population concentrates in the $|m=3\rangle$  state during optical pumping. $\xi_1$ decreases but $\xi_2$ increases and approaches unity along with the optical pumping time. At longer pumping times, the overall $\xi_c^R$ approaches a value of less than unity.

For an isotropic Zeeman population distribution at the zero pumping time, the finite-bandwidth factor ($\xi_1$) is equal to unity since the last two terms in the bracket in Eq. (\ref{xi_i_c2}) cancel each other out. $\xi_c^R$ is only determined by the coherence mismatch factor $\xi_2$. This is true for both the $\sigma^+ \rightarrow \sigma^-$ and the $\sigma^+ \rightarrow \sigma^-$ conversion system and for an optical depth of any value. This is why the four curves in Fig. \ref{xi_i_pump} all merge to the same value at zero pumping time. 

Next we examine the relation between $\xi_c^R$ and $\eta$ for four different population distributions, as shown in Fig.\ref{xi_i_eta}. In case (d) in Fig. \ref{xi_i_eta} with an isotropic population distribution, $\xi_c^R$ is independent of $\eta$ because of the cancellation of the finite bandwidth effect, as mentioned above. In cases (c) and (d), the ground-state coherence mismatch factor $\xi_2$ still surpasses the finite bandwidth factor $\xi_1$ such that $\xi_c^R$ is not greater than unity for all shown values of $\eta$. With a higher concentrated population as in case (a) and at a large enough $(\eta >6)$, the $\xi_1$ factor dominates over $\xi_2$ such that $\xi_c^R$ is larger than unity. It can be seen that increasing $\eta$ strengthens the effect of finite bandwidth. 

\begin{figure}[t]
\centering
\includegraphics[width=0.52\textwidth]{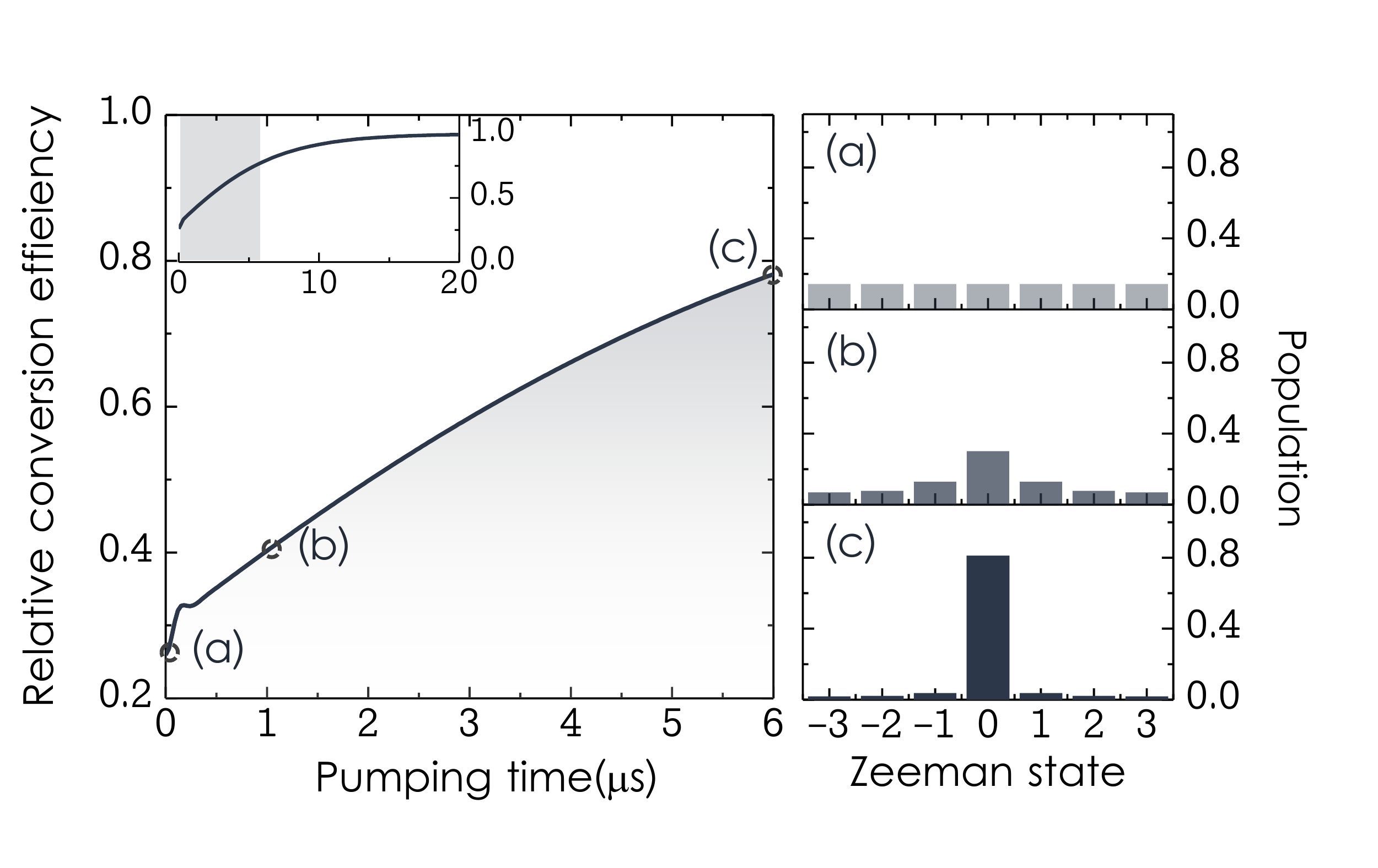}
\caption{In the process of pumping to the $|m=0\rangle$ state, the relative conversion efficiency is exactly equal to the ground-state coherence mismatch factor $\xi_2$. As the population gathers in the $m=0$ state, $\xi_c^I$ gradually approaches unity, as shown in the inset. Figure (a)-(c) represent the population distributions at pumping times of 0$\mu$s, 1$\mu$s and 6$\mu$s, respectively.}
\label{xi_i_pump_m0}
\end{figure}

Case (d) in Fig.\ref{xi_i_eta} offers a possible solution for elimination of the finite bandwidth effect. The finite bandwidth effect does not appear when the effective optical depth of the reading or writing process is the same. The cases with symmetric population distribution with respect to the $|m=0\rangle$ state all share this feature. Therefore, consider the optical pumping which drives the $\pi$ transition. In this condition, the Zeeman population is symmetrically distributed w.r.t. the $|m=0\rangle$ state. For long pumping times, all the population concentrates towards the single Zeeman state $|m=0\rangle$. For symmetric population, $\xi_1=1$ and only the ground-state coherence mismatch factor $\xi_2$ affects $\xi_c^R$ and thus $\xi_c^R=\xi_2$.  Since the bandwidth effect does not appear in this case, $\xi_c^R$ is not greater than unity, as shown in Fig.\ref{xi_i_pump_m0}. For the case where the entire population is in the $|m=0\rangle$ state (long pumping times as shown in inset to Fig. \ref{xi_i_pump_m0}), $\xi_c^R$ is equal to unity and is free from both the finite bandwidth and the coherence mismatch factors.

\section{\label{sec6}Conclusion}
In conclusion, we carried out a detailed study on an EIT-memory-based light field converter with degenerate Zeeman states. We discuss the process of reading and writing in the conversion system and derive an approximate analytical solution for the converted field, which clarifies that the effect of the finite EIT bandwidth effect and the ground-state coherence mismatch are the two limiting factors for the conversion efficiency. We discuss how these two factors affect the overall conversion efficiency for various population distributions among the Zeeman states. Our work provides essential physical insights and quantitative knowledge for the application of EIT-memory-based light conversion with degenerate Zeeman states.  

\section*{ACKNOWLEDGMENT}
This work was supported by the Ministry of Science and Technology of Taiwan under grant numbers 106-2119-M-001-002, 107-2112-M-001-003, 107-2745-M-007-001, 108-2639-M-007-001-ASP and 108-2112-M-001-030-MY3. We also thank the Center for Quantum Technology and NCTS ECP1 of Taiwan for their supports, as well as Ite A. Yu and Yong-Fan Chen for their helpful discussions. 

\section*{Appendix}
The numerical simulation of optical pumping in cesium with a pump beam that drives $\left|6^2S_{1/2},F=3\right\rangle \rightarrow \left|6^2P_{3/2},F'=3\right\rangle$ is shown in detail. The seven Zeeman states in the  $\left|6^2S_{1/2},F=3\right\rangle$ ground states and the seven Zeeman states in the excited states in $\left|6^2P_{3/2},F'=3\right\rangle$ are included in the calculation. The Hamiltonian of the atom-photon interaction due to the optical pumping beam is given by
\begin{equation}
\begin{aligned}
\hat{H}=-\frac{1}{2}\hbar(&\sum_{j=-3}^{2}b_{r,j}\Omega_{r}\hat{\sigma}_{j,j+1}\sum_{j=-3}^{3}b_{\pi,j}\Omega_{\pi}\hat{\sigma}_{j,j}\\
+&\sum_{j=-2}^{3}b_{l,j}\Omega_l\hat{\sigma}_{j,j-1}+H.C.),
\label{H_pump}
\end{aligned}
\end{equation}
where $b_{(r,\pi,l),j}$ denotes the CG-coefficient for the $\sigma^+$, $\pi$ and  $\sigma^-$ transition from the state $\left|F=3,j\right\rangle$, respectively. The operator $\hat{\sigma}_{m,n}$ represents the flip operators that describe the transition from the state $\left|F=3,m\right\rangle$ to $\left|F'=3,n\right\rangle$. $\Omega_{r,\pi,l}$ are the Rabi frequencies of the pumping field component for the transition of $\sigma^+$, $\pi$ and  $\sigma^-$, respectively. The equations of motion for the atomic coherence and population are given by the optical Bloch equations:
\begin{equation}
\begin{aligned}
\partial_t\langle\hat{\sigma}_{m,n}\rangle=\frac{i}{\hbar}\left\langle\left[\hat{H},\hat{\sigma}_{m,n}\right]\right\rangle-\Gamma_{m,n}\langle\hat{\sigma}_{m,n}\rangle,
\label{OBE_pump}
\end{aligned}
\end{equation}
where $\Gamma_{m,n}$ is the decay rates of $\sigma_{m,n}$. Although in the actual cesium atoms, the population in the excited state $\left|6^2P_{3/2},F'=3\right\rangle$ can relax to the $\left|6^2S_{1/2},F=4\right\rangle$ ground state. To simplify the calculation, we renormalize the spontaneous decay rate such that the excited-state population can relax to the $\left|6^2S_{1/2},F=3\right\rangle$ ground state only and neglect the nine Zeeman sublevels of the $\left|6^2S_{1/2},F=4\right\rangle$ state. We emphasize that although this optical pumping simulation may not be physically precise, it captures the main feature of Zeeman optical pumping. This model offers the dynamic population distributions among the Zeeman sublevels of the $\left|6^2S_{1/2},F=3\right\rangle$ state for a discussion of their dependence on the conversion efficiency. By combining Eqs.(\ref{H_pump}) and (\ref{OBE_pump}), we can calculate the dynamics of the population distribution. For the case in Fig.\ref{pumping}, we set $\Omega_{\pi,l}=0$ and $\Omega_{r}=1.2\Gamma$. In fig.\ref{xi_i_pump_m0}, the conditions for the pumping field are $\Omega_{\pi}=1.2\Gamma$ and $\Omega_{r,l}=0$.




\input{ref.bbl}

\end{document}

%% file: ref.bbl
%